\documentclass{PoS}

\usepackage{epsfig}

\title{Confinement and Green functions in Landau-gauge QCD}
\ShortTitle{Confinement and Green functions in Landau-gauge QCD}

\author{\speaker{Reinhard Alkofer}\\
Institut f\"ur Physik,
Karl-Franzens-Universit\"at, 
Universit\"atsplatz 5,
A-8010 Graz, Austria\\
E-mail: \email{reinhard.alkofer@uni-graz.at}}
\author{Christian S. Fischer\\
        Institut f\"ur Physik, TU Darmstadt,
	Schlossgartenstr. 9,
	D-64289 Darmstadt, Germany}
\author{Markus Q.\ Huber\\
Institut f\"ur Physik,
Karl-Franzens-Universit\"at, 
Universit\"atsplatz 5,
A-8010 Graz, Austria}	
\author{Felipe J. Llanes-Estrada\\
Universidad Complutense de
Madrid, Depto. F\'{\i}sica Te\'orica I. 28040 Madrid, Spain}
\author{Kai Schwenzer\\
Institut f\"ur Physik,
Karl-Franzens-Universit\"at, 
Universit\"atsplatz 5,
A-8010 Graz, Austria}

\abstract{
In a functional approach to  QCD the infrared behaviour of Landau gauge
Green functions is investigated. Positivity violation for, and thus
confinement of, gluons is demonstrated, and the analytic structure of 
the gluon propagator is studied.
Quark confinement is related to an infrared divergence of the quark-gluon
vertex. In the latter various components are dynamically induced due to the
spontaneous breaking of chiral symmetry. As a result an infrared finite
running coupling in the Yang-Mills sector is derived whereas the running
coupling related to the quark-gluon vertex is infrared divergent.
Based on a truncation for the
quark-gluon vertex Dyson-Schwinger equation, which respects the analytically
determined infrared behavior, numerical results for the coupled system of the
quark propagator and vertex Dyson-Schwinger equations  are presented. 
The resulting quark mass function as well as the vertex function show only a
very weak dependence on the current quark mass in the deep infrared. 
From this
we infer by an analysis of the quark-quark scattering kernel a linearly rising
quark potential with an almost mass independent string tension in the case of
broken chiral symmetry.}

\FullConference{8th Conference Quark Confinement and the Hadron Spectrum\\
                 September 1-6, 2008\\
                 Mainz, Germany}

\begin{document}

\section{Infrared behaviour of Landau gaugeYang-Mills theory \label{IRYM}}
 
Confinement, the $U_A(1)$ anomaly, and dynamical chiral symmetry breaking are
supposed to be properties of infrared QCD. Despite the progress achieved in our
understanding of these phenomena the underlying mechanisms as well as possible
interrelations between them are not yet uncovered. This talk aims at summarizing
what we can learn from QCD Green functions in the Landau gauge about these
topics.

Explaining quark confinement, and hereby especially extracting the linearly
rising static quark-antiquark potential and relating its properties to QCD
degrees of freedom, has been the main objective of many quite different
studies. In  ref.\ \cite{Alkofer:2006fu} some of these
pictures where quark confinement is related to
\begin{itemize}
\addtolength{\itemsep}{-3mm}
\item the condensation of chromomagnetic monopoles (e.g.\
\cite{Mandelstam,Pisa}),
\item the percolation of center vortices(e.g.\ \cite{Greensite}),
\item the AdS$_5$ / QCD correspondence (e.g.\ \cite{Maldacena:1998im}),
\item the Gribov-Zwanziger scenario in Coulomb gauge (e.g.\  
\cite{Gribov,Dan1}), or
\item the infrared behaviour of Landau gauge Green functions
\cite{Alkofer:2000wg,Fischer:2006ub,Alkofer:2006jf},
\end{itemize}
have been briefly reviewed.
These explanations for confinement are seemingly
different but there are surprising relations between them which are not yet
understood. Given the current status one has to note that these pictures are
definitely not mutually exclusive but simply reveal only different aspects of
the confinement phenomenon. A similar statement on the relation between 
seemingly different scenarios is definitely true when considering the $U_A(1)$ 
anomaly, see {\it e.g.\/} the discussion in ref.\ \cite{Alkofer:2008et}.
This should be kept in mind when in the following
the results based on investigations of Landau gauge Green functions are
presented.

Employing Green functions dynamical chiral symmetry breaking reflects itself
directly in terms of the quark propagator, see refs.\
\cite{Alkofer:2000wg,Fischer:2006ub,Miransky:1985ib,Pennington:1998cj}  and
references therein. However, as we have demonstrated recently this picture is
incomplete as it does not take into account the effects of dynamical chiral
symmetry breaking on higher $n$-point functions 
\cite{Alkofer:2006gz,Alkofer:2007qf,Alkofer:2008tt}. 
This issue will be discussed in detail in
this talk. To derive it one needs first some insight into the infrared behaviour
of gluons and ghosts.

\subsection{Infrared exponents of gluons and ghosts}
\begin{figure}[th]
\centerline{\psfig{file=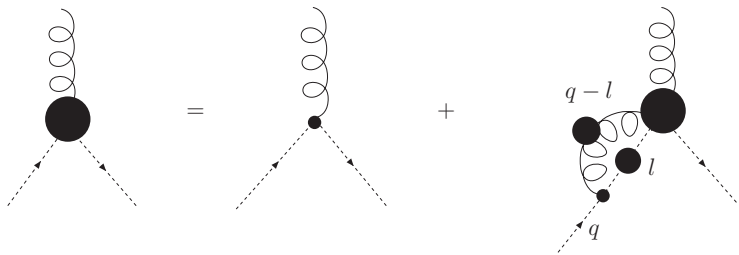,width=100mm}}
\vspace*{-8pt}
\caption{The Dyson-Schwinger equation for the ghost-gluon vertex.
\label{GhGlDSE}}
\end{figure}
The possibly best starting point for the investigation of the infrared behaviour
of gluons and ghosts is given by the Dyson-Schwinger 
equation for the ghost-gluon vertex function, see fig.\ \ref{GhGlDSE}.
(NB: A MATHEMATICA package to derive Dyson-Schwinger equations can be found in
\cite{Alkofer:2008nt}.)
In the Landau gauge the gluon propagator is transverse, and therefore one can
employ the relation
\begin{equation}
l_\mu D_{\mu \nu}(l-q) = q_\mu D_{\mu \nu}(l-q) \, ,
\end{equation}
to conclude that the ghost-gluon vertex stays
finite when the outgoing ghost momentum vanishes, {\it i.e.} when $q_\mu
\rightarrow 0$~\cite{Taylor:1971ff}. This argument is valid to all orders in
perturbation theory, a truely non-perturbative justification of the 
infrared finiteness of this vertex has been given in refs.\
\cite{Lerche:2002ep,Cucchieri:2004sq,Schleifenbaum:2004id}.

\begin{figure}[th]
\centerline{\psfig{file=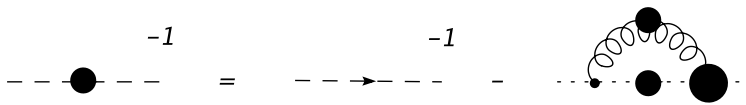,width=100mm}}
\vspace*{-8pt}
\caption{The Dyson-Schwinger equation for the ghost propagator.
\label{GhDSE}}
\end{figure}

Knowing this property of the ghost-gluon vertex
the Dyson-Schwinger equation for the ghost propagator,
see fig.\ \ref{GhDSE}, can be analysed.
The only unknown functions in the deep infrared are the gluon and the ghost 
propagators, parameterized in the Euclidean domain as
\begin{eqnarray}
        D_{\mu\nu}(k) = \frac{Z(k^2)}{k^2} \, \left( \delta_{\mu\nu} -
        \frac{k_\mu k_\nu}{k^2} \right)  \; ,\quad
        D_G(k)  &=& - \frac{G(k^2)}{k^2}          \;.
\end{eqnarray}
In Landau gauge these propagators are best described by
two invariant functions, $Z(k^2)$ and $G(k^2)$, respectively.
As solutions of renormalized equations,  these functions 
depend also on the renormalization  scale $\mu$. Furthermore, assuming that 
QCD Green functions can be expanded in asymptotic series, the integral in the
ghost Dyson--Schwinger equation can be split up in three pieces: an infrared
integral, an ultraviolet integral,  and an expression for the ghost wave
function renormalization. Hereby it is the resulting equation for the latter
quantity which allows one to extract definite information \cite{Watson:2001yv}
without using any truncation or ansatz.

Recently it became clear that there are two distinct types of solutions, see
ref.\ \cite{Fischer:2008uz} and references therein. In this talk we will focus
on the so-called scaling solution where  the infrared behaviour of the gluon and
ghost propagators is given by power laws. The corresponding exponents are
uniquely related such that the gluon exponent is minus two times the ghost
exponent \cite{vonSmekal:1997is}. As we will see later on this implies an
infrared fixed point for the corresponding running coupling. The signs of the
exponents are such that the gluon propagator is infrared suppressed as compared
to the one for a free particle, the ghost propagator is infrared enhanced.

\begin{figure}[th]
\centerline{\psfig{file=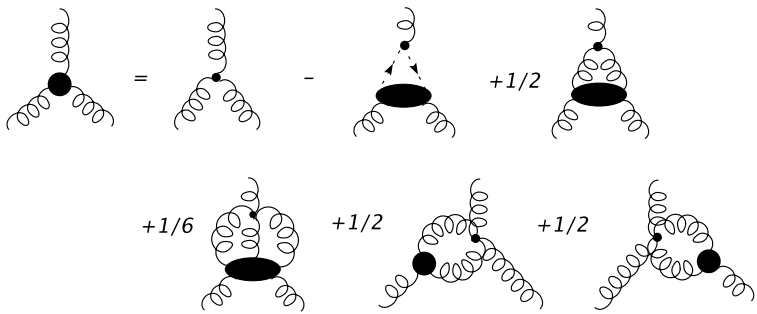,width=100mm}}
\vspace*{-8pt}
\caption{The Dyson-Schwinger equation for the 3-gluon vertex.
\label{3GDSE}}
\end{figure}
\begin{figure}[th]
\centerline{\psfig{file=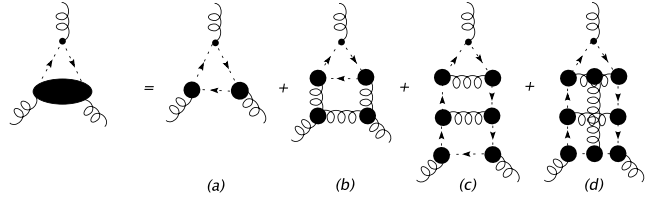,width=90mm}}
\vspace*{-8pt}
\caption{An example for the skeleton expansion of the
3-gluon vertex.
\label{3Gskel}}
\end{figure}
\begin{figure}[th]
\centerline{\psfig{file=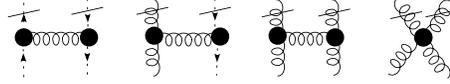,width=60mm}}
\vspace*{-8pt}
\caption{Insertions to reconstruct higher orders in the skeleton expansion.
\label{3GskelIn}}
\end{figure}

Using this infrared power laws for the Yang-Mills propagators  one can 
infer the infrared behaviour of higher $n$-point functions. To this
end the  $n$-point Dyson-Schwinger equations have been studied in a
skeleton expansion, {\it i.e.\/} a loop expansion using dressed propagators and
vertices. Furthermore, an asymptotic expansion has been applied to all primitively
divergent Green functions \cite{Alkofer:2004it}.
As an example consider the Dyson-Schwinger equation for the 3-gluon vertex which
is diagrammatically represented in fig.\ \ref{3GDSE}.
Its skeleton expansion, see fig.\ \ref{3Gskel}, can be constructed
via the insertions given in fig.\ \ref{3GskelIn}.
These insertions have vanishing infrared anomalous dimensions which implies
that the resulting higher order terms feature the same infrared scaling.
Based on this
the following general infrared behaviour for one-particle irreducible
Green functions with $2n$ external ghost legs and $m$ external
gluon legs can be derived \cite{Alkofer:2004it,Huber:2007kc}:
\begin{equation}
\Gamma^{n,m}(p^2) \sim (p^2)^{(n-m)\kappa + (1-n)(d/2-2)}  \label{IRsolution}
\end{equation}
where $\kappa$ is one yet undetermined parameter, and $d$ is the
space-time dimension.

Here two remarks are in order: First, exploiting
Dyson-Schwinger {\bf and} Exact Renormalization Group Equations
one can show that this infrared solution is unique \cite{Fischer:2006vf}.
Second, there are additional divergences when only some of the momenta of the
$n$-point functions are vanishing \cite{Alkofer:2008jy}.

\subsection{Infrared fixed point of the running coupling in the Yang-Mills
sector}

The infrared solutions (\ref{IRsolution}) include
\begin{equation}
{ G(p^2) \sim (p^2)^{-\kappa}}\;,
\quad { Z(p^2) \sim (p^2)^{2\kappa}}            
\quad {  \Gamma^{3g}(p^2) \sim (p^2)^{-3\kappa}} \;,
\quad { \Gamma^{4g}(p^2) \sim (p^2)^{-4\kappa}} \;.
\end{equation}
This implies that the running couplings related to these
vertex functions possess an infrared fixed point:
\begin{eqnarray}
\displaystyle \alpha^{gh-gl}(p^2) &=&
\alpha_\mu \, { G^2(p^2)} \, { Z(p^2)}
\sim \frac{const_{gh-gl}}{N_c} ,
\quad
\displaystyle \alpha^{3g}(p^2) =
\alpha_\mu \, { [\Gamma^{3g}(p^2)]^2} \, { Z^3(p^2)}
\sim \frac{const_{3g}}{N_c} ,
\nonumber \\
\displaystyle \alpha^{4g}(p^2) &=&
\alpha_\mu \, {  \Gamma^{4g}(p^2)} \, { Z^2(p^2)}
\sim \frac{const_{4g}}{N_c}.
\end{eqnarray}
The infrared value of the coupling related to the ghost-gluon
vertex can be computed to be \cite{Lerche:2002ep,Fischer:2002hn}:
\begin{equation}
\alpha^{gh-gl}(0)=\frac{4 \pi}{6N_c}
\frac{\Gamma(3-2\kappa)\Gamma(3+\kappa)\Gamma(1+\kappa)}{\Gamma^2(2-\kappa)
\Gamma(2\kappa)} .
\end{equation}
This yields $\alpha^{gh-gl}(0)=2.972$ for $N_c=3$ and $\kappa = (93 -
\sqrt{1201})/{98} \simeq 0.595353 $, which is the value obtained with a bare
ghost-gluon vertex.

\subsection{Gluon confinement by positivity violation for the gluon propagator}

Positivity violation for a propagator entails that the corresponding field is
not related to an asymptotic state and thus to a particle. In the cases of
gluons one can infer gluon confinement from such a scenario, see {\it e.g.\/} 
\cite{vonSmekal:2000pz} and references therein. Therefore
positivity violation of the propagator of transverse gluons has
been for a long time a conjecture which has been convincingly verified by now,
\cite{Alkofer:2003jj,Bowman:2007du} as well as references
therein. The basic feature is hereby the infrared suppression of transverse 
gluons caused by the infrared enhancement of ghosts. 
Being related to the confinement of tranverse gluons
\cite{vonSmekal:2000pz} it is certainly worth to have a closer look at the
underlying analytic structure of the gluon propagator.

As the infrared exponent $\kappa$ in the infrared power laws is an irrational
number this implies already that the gluon propagator possesses a cut on the
negative real $p^2$ axis. It is possible to fit the solution for the gluon
propagator accurately without introducing further singularities in the
complex $p^2$ plane \cite{Alkofer:2003jj}:
\begin{equation}
Z_{\rm fit}(p^2) = w \left(\frac{p^2}{\Lambda^2_{\tt QCD}+p^2}\right)^{2 \kappa}
 \left( \alpha_{\rm fit}(p^2) \right)^{-\gamma} .
 \label{fitII}
\end{equation}
Hereby $w$ is a normalization parameter, and
$\gamma = (-13 N_c + 4 N_f)/(22 N_c - 4 N_f)$
is the one-loop value for
the anomalous dimension of the gluon propagator.
The running coupling is expressed as \cite{Fischer:2003rp}:
\begin{eqnarray}
\alpha_{\rm fit}(p^2) &=& \frac{\alpha_S(0)}{1+p^2/\Lambda^2_{\tt QCD}}
+\frac{4 \pi}{\beta_0} \frac{p^2}{\Lambda^2_{\tt QCD}+p^2}
\left(\frac{1}{\ln(p^2/\Lambda^2_{\tt QCD})}
- \frac{1}{p^2/\Lambda_{\tt QCD}^2 -1}\right) 
\end{eqnarray}
with $\beta_0=(11N_c-2N_f)/3$.
Note that the gluon propagator (\ref{fitII}) possesses an important property:
{\em Wick rotation is directly possible!}

\subsection{Summary on Yang-Mills sector}

To summarize this section we note that the here discussed scaling solution for
Green functions in an Yang-Mills theory implies that the gluon propagator
vanishes on the light cone, and $n$-point gluon vertex functions diverge
on the light cone. Therefore an attempt to kick a gluon free, {\it i.e.} to
produce a real gluon, immediately results in production of infinitely many
virtual soft gluons to produce perfect color charge screening. Positivity
violation (which implies BRST quartet cancelation \cite{vonSmekal:2000pz}) 
guarantees that this screening is total, or phrased otherwise, that gluons are
confined. 

\section{Dynamically induced scalar quark confinement}

\subsection{Self-consistency between 
the quark propagator and the quark-gluon vertex}

\begin{figure}[th] 
\centerline{\psfig{file=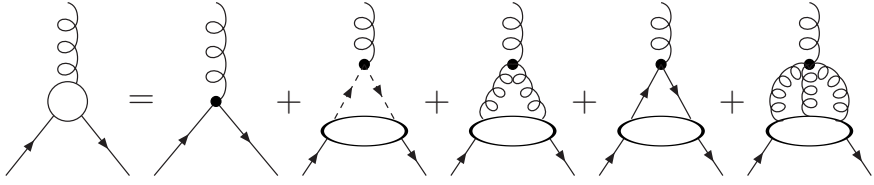,width=120mm}}
\vspace*{-8pt}
\caption{The
 Dyson-Schwinger equation  for the quark-gluon vertex.
\label{QGV}}
\end{figure}

\begin{figure}[th]
\centerline{\psfig{file=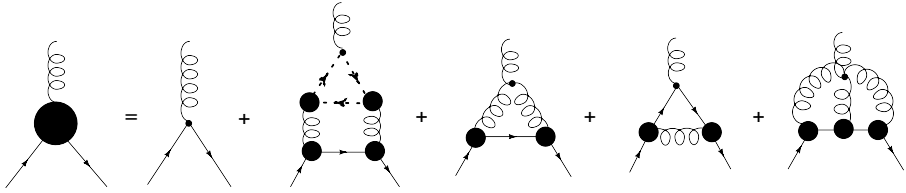,width=120mm}}
\vspace*{-8pt}
\caption{The skeleton expansion   for the quark-gluon vertex.
\label{QGV-skel}}
\end{figure}

As detailed above Landau gauge Green functions provide a consistent picture for 
gluon confinement. However, due to the infrared suppression of the gluon
propagator quark confinement seems even  more unexplainable.  To
proceed it turns out to be necessary to study the Dyson-Schwinger equation
for the quark propagator together with the one for the quark-gluon
vertex \cite{Alkofer:2000wg,Fischer:2006ub,
Alkofer:2006jf,Alkofer:2008tt,Schwenzer:2008vt,Fischer:2003rp}.
Therefore a detailed study of this three-point function, and especially its
infrared behaviour, is mandatory. Its Dyson-Schwinger equation is
diagrammatically depicted in fig.\ \ref{QGV}, its skeleton expansion in
fig.\ \ref{QGV-skel}.
Hereby a drastic difference of the quarks as compared
to Yang-Mills fields has to be taken into account: They possess a mass,
and in addition one has to take into account that dynamical chiral
symmetry breaking and thus dynamical mass generation may (and will) occur.

The infrared analysis of the Yang-Mills theory has been generalized to full,
although quenched, QCD in ref.\ \cite{Alkofer:2008tt}.  
Let us first consider a limit where the masses of the valence quarks are large,
{\it i.e.\/}  $m > \Lambda_{\tt QCD}$.   The remaining scales below 
$\Lambda_{\tt QCD}$ 
are those of the external momenta of the propagators and vertex functions. 
The Dyson-Schwinger equations are then employed 
to determine the selfconsistent solutions  in
terms of powers of the small external momentum scale $p^2 \ll \Lambda_{\tt
QCD}$. The equations which have to be considered in addition to the ones of
Yang-Mills theory are the ones for the quark propagator and the quark-gluon
vertex which in turn have to be solved self-consistently.

\subsection{The quark-gluon vertex, chiral symmetry, and quark confinement}

The fully renormalized quark-gluon vertex $\Gamma_\mu$ consists of up to 
twelve linearly independent Dirac tensors. Half of these vanish in case chiral
symmetry would be realized in the Wigner-Weyl mode, {\it i.e.\/}  these
tensor structures can only be non-vanishing either if chiral symmetry is explicitely
broken by current masses and/or chiral symmetry is realized in Nambu-Goldstone
mode ({\it i.e.} spontaneously broken). From a solution of the Dyson-Schwinger
equations we infer that  these {\em ``scalar''} structures are, in the chiral
limit, generated non-perturbatively together with the dynamical quark mass
function in a self-consistent fashion. Thus dynamical chiral symmetry breaking
reflects itself not only in the propagator but also in the quark-gluon vertex.

\begin{figure}[th] 
\centerline{\psfig{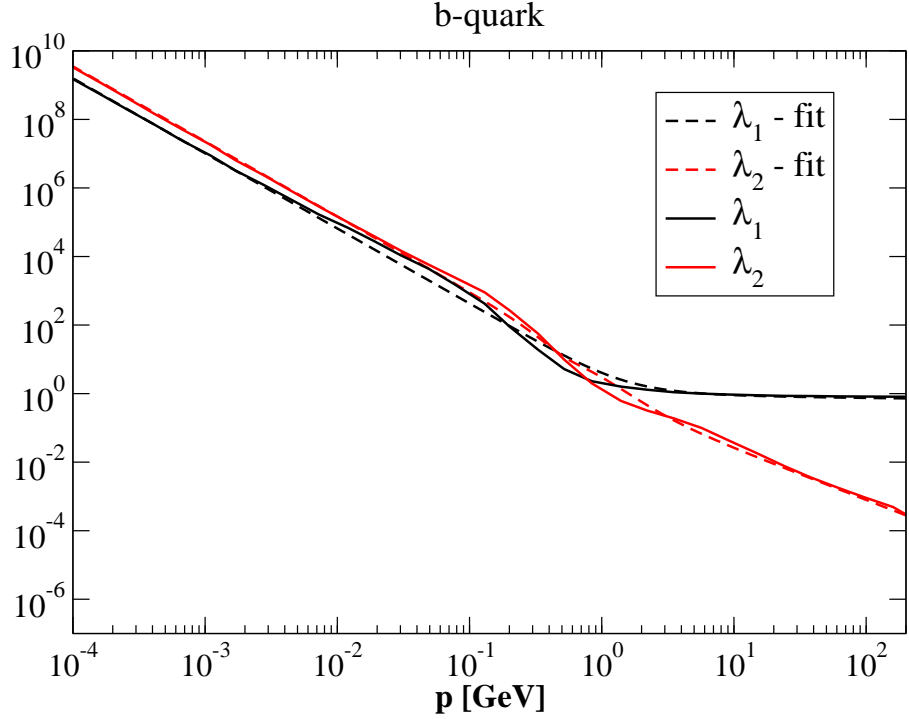}}
\vspace*{-8pt}
\caption{The leading vector component ${\lambda}_1$ and a scalar component
${\lambda}_2$ of the quark-gluon vertex for a current mass corresponding to a 
$b$ quark. Adapted from ref.\ \cite{Alkofer:2008tt}
\label{ResQGV}}
\end{figure}

Performing an infrared analysis one obtains
an infrared divergent solution for the quark-gluon vertex
such that  Dirac vector and {\em ``scalar''}
components of this  vertex are infrared divergent with exponent $-\kappa -
\frac 1 2$ when either all momenta or when only the gluon momenta
vanish \cite{Alkofer:2006gz,Alkofer:2008tt}. 
A numerical solution of a truncated set of
Dyson-Schwinger equations confirms this infrared behavior, see
fig.~\ref{ResQGV}. The
driving pieces of this solution are the scalar Dirac amplitudes of the
quark-gluon vertex and the scalar part of the quark propagator. Both pieces are
only present when chiral symmetry is broken, either explicitely or dynamically.
As can be seen from fig.~\ref{ResQGV} the function $\lambda_1$ multiplying the
tree-level tensor structure $ \gamma_\mu$ is only the leading one in the
ultraviolet, in the infrared the scalar component ${\lambda}_2$ is even larger
than the vector component ${\lambda}_1$.

The quark-gluon vertex may also serve to extract a running coupling. Using
\begin{equation}
{\Gamma^{qg}(p^2) \sim (p^2)^{-1/2-\kappa}} \,, \ \
{ Z_f(p^2) \sim const}\,, \ \ { Z(p^2) \sim (p^2)^{2\kappa}} 
\end{equation}
one obtains
\begin{equation}
\alpha^{qg}(p^2) = \alpha_\mu \,
{ [\Gamma^{qg}(p^2)]^2} \, { [Z_f(p^2)]^2}\,
{ Z(p^2)} \sim  \frac{const_{qg}}{N_c} \frac{1}{p^2} ,
\label{aq}
\end{equation}
and thus a coupling which is  singular in the infrared contrary to the
couplings  from the Yang-Mills vertices.

To determine whether this already relates to quark confinement the anomalous
infrared exponent of the four-quark function is determined.  
The static quark potential can be obtained
from this four-quark one-particle irreducible Green function, which, including
the canonical dimensions, behaves like $(p^2)^{-2}$for $p^2\to0$ due to the
singularity of the quark-gluon vertex for vanishing gluon momentum.
Using a well-known relation for a function $F\propto (p^2)^{-2}$
yields 
\begin{equation}
V({\bf r}) = \int \frac{d^3p}{(2\pi)^3}  F(p^0=0,{\bf p})  e^{i {\bf p r}}
\ \ \sim \ \ |{\bf r} |
\end{equation}
for the static quark-antiquark potential $V({\bf r})$.
At this point one notes that, given the infrared divergence of the
quark-gluon vertex as found in the scaling solution of the coupled system of
Dyson-Schwinger equations, the vertex overcompensates the infared suppression
of the gluon propagator such that one obtains a linearly rising potential.
The surprising fact is that this potential is dynamically induced and 
in the infrared dominantly scalar.

To provide further understanding for the here found relation between chiral 
symmetry breaking and quark confinement one may keep chiral symmetry 
artificially  in Wigner-Weyl mode, {\it i.e.} in the chiral limit one forces
the quark mass term as well as the ``scalar'' terms in the quark-gluon vertex
to be zero. The result of such a procedure is that 
then the running coupling from the quark-gluon vertex is no longer
diverging but goes to a fixed point in the infrared similar to the couplings
from the Yang-Mills vertices. Correspondingly, one obtains 
a $1/r$ behaviour of the static quark potential.
The ``enforced'' restoration of chiral symmetry is therefore directly linked 
with the disappearance of  quark confinement. The infared properties of the
quark-gluon vertex in the ``unforced'' solution thus constitute a novel
mechanism that {\em directly links chiral symmetry breaking with quark
confinement.}

\subsection{Mass dependence}

\begin{figure}[th] 
\centerline{\psfig{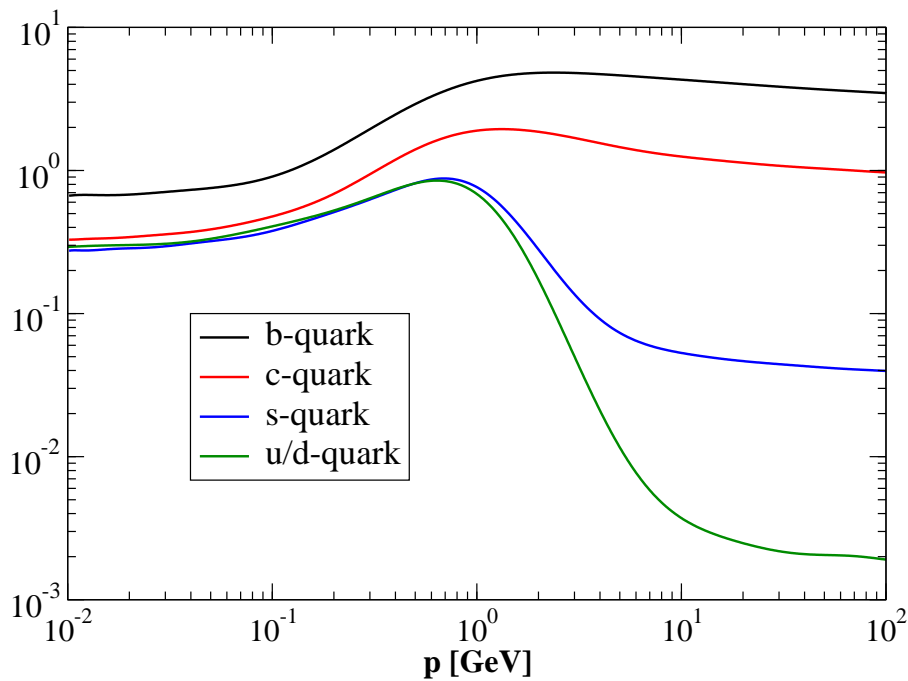}}
\vspace*{-8pt}
\caption{The quark mass functions for several values of current masses.
 Adapted from ref.\ \cite{Alkofer:2008tt}
\label{ResQM}}
\end{figure}

As mentioned above, based on a truncation for the
quark-gluon vertex Dyson-Schwinger equation which respects the analytically
determined infrared behavior, numerical results for the coupled system of the
quark propagator and vertex Dyson-Schwinger equations have been obtained 
\cite{Alkofer:2008tt}. 
In fig.~\ref{ResQM} the resulting quark mass function is presented for the
values of current masses related to $u/d$, $s$, $c$ and $b$ quarks. 
These quark mass functions as well as the vertex functions  show only a
very weak dependence on the current quark mass in the deep infrared.
First of all, the infrared exponent for all vertex functions are mass
independent. Second, comparing the coefficients of these power laws, given in 
table \ref{table}, we see a much smaller dependence on the quark mass than 
could have been anticipated from a naive analysis. As the static potential
will be
build up by the infrared divergent parts of the quark-gluon vertex we thus
predict that the string tension is almost mass-independent. Provided the
Coulombic part of potential has still a large effect in
the bottomonium spectrum this is in agreement with experimental splittings of
charmonia and bottonia systems.
\begin{table}[t]
\begin{center}
\begin{tabular}{c|c|c|c|c|}
                                & $u/d$ & $s$  & $c$  & $b$   \\\hline
$m(\mu^2) [MeV]              $  &   3   &  50  & 1200 & 4200  \\\hline
$M(0)     [MeV]              $  &  270  & 270  &  320 & 650   \\\hline
$\lambda_1 [GeV^{1/2+\kappa}]$  &  3.95 & 3.60 & 2.00 & 2.73  \\\hline
$\lambda_2 [GeV^{1/2+\kappa}]$  &  8.70 & 7.97 & 4.44 & 6.08
\end{tabular}
\end{center}
\caption{The infrared coefficients for the dressing functions $\lambda_{1,2}$ for
different current quark masses $m(\mu^2)$ (at the renormalization point 
$\mu^2=170
\,\mbox{GeV}$). Also given are the values of the mass function at zero
momentum.\label{table}}
\end{table}

Thus our analysis provides (at least) two surprises: First, the longstanding 
discussion on whether confinement is of a vector or of a scalar nature is 
oversimplified. Dynamically generated structures  of vector, scalar and tensor
types in the quark-gluon vertex lead to a rich structure of the quark-antiquark
interaction. Second, the mass dependence of the string tension does not seem to
follow any easy rule, {\it cf.} also recent improved lattice data
\cite{Koma:2007jq}. These and other details of the confining interaction 
clearly deserve more investigations.

\subsection{Summary and outlook on quark sector}

The scaling solution for Landau-gauge QCD Green functions has far reaching
consequences for the infrared behaviour of quarks. The resulting dynamical
chiral symmetry breaking and the related dynamical mass generation lead to an
infrared trivial quark propagator. But dynamical chiral symmetry breaking  also
occurs in the quark-gluon vertex. Hereby several components, and especially also
the ``scalar'' ones,  diverge on the quark ``mass'' shell.

An attempt to kick a quark free, {\it i.e.} to produce a real quark, immediately
results in production of infinitely many virtual soft gluons. Hereby these
gluons do not only couple vector-like but also scalar-like! And the vertex
function diverges such that effectively a linearly rising potential is produced.
One obtains therefore infrared slavery and quark confinement.

Although these results are in itself quite surprising and encouraging they
provide only the starting point for an understanding of quark confinement in
functional approaches. Questions which will be investigated further include:
\begin{itemize}
\addtolength{\itemsep}{-3mm}
\item How does the formation of a string between colour sources reflect itself
in Green functions?
\item Can we learn about the properties of the confining field configurations
from Green functions?
\item How does all this relate to other (partially) successful approaches to
understand the phenomenon of quark confinement? 
\end{itemize}

\section*{Acknowledgements}

R.A.\ thanks the organisers of {\it Confinement 8\/}
for all their efforts which made this extraordinary conference possible.
We are grateful to A.~Cucchieri, A.~Maas, T.~Mendez,  J.~Pawlowski,
 and L.~v.~Smekal for interesting discussions. \\
R.A.\ was supported by the FWF grant P20592-N16,
C.S.~F.\  the  Helmholtz-University Young
  Investigator Grant No. VH-NG-332, and K.~S.\
by the FWF Lise-Meitner grant M979-N16.

\end{document}